\documentclass[prd,10pt,aps,twocolumn,amsmath,amssymb,superscriptaddress,longbibliography,floatfix]{revtex4-1}

\usepackage[colorlinks=true, linkcolor=violet, citecolor=teal, urlcolor=blue]{hyperref}

\usepackage{cancel}
\usepackage[normalem]{ulem}

\usepackage{nicefrac}
\usepackage{graphicx}
\usepackage{float}
\usepackage{combelow}
\usepackage{dcolumn}		%decimal point alignment in Tables, use 'd'
\usepackage[dvipsnames]{xcolor}
\newcommand{\replaceR}[2]{{{\color{BrickRed}{#1}}{\color{NavyBlue}{\ifmmode\text{\sout{\ensuremath{#2}}}\else\sout{#2}\fi}}}}

\newcommand{\replaceB}[2]{{{\color{RedViolet}{#1}}{\color{BlueViolet}{\ifmmode\text{\sout{\ensuremath{#2}}}\else\sout{#2}\fi}}}}
\newcommand{\vb}[2]{{{\color{RedViolet}{#1}}{\color{BlueViolet}{\ifmmode\text{\sout{\ensuremath{#2}}}\else\sout{#2}\fi}}}}

\newcommand{\replaceC}[2]{{{\color{RedOrange}{#1}}{\color{OliveGreen}{\ifmmode\text{\sout{\ensuremath{#2}}}\else\sout{#2}\fi}}}}

\newcommand{\bs}{\boldsymbol}

\allowdisplaybreaks

\begin{document}

\title{Acceleration as refrigeration: \\
Acceleration-induced spontaneous symmetry breaking in thermal medium}

\author{Maxim~N.~Chernodub}
\email{maxim.chernodub@univ-tours.fr}
\affiliation{Institut Denis Poisson, CNRS - UMR 7013, Universit\'e de Tours, 37200 France}
\affiliation{Department of Physics, West University of Timi\cb{s}oara,  Bd.~Vasile P\^arvan 4, Timi\cb{s}oara}

\date{\today}

\begin{abstract}
We argue that a uniform acceleration of matter produces an effect similar to cooling, thus leading, in particular, to the enhancement effect of spontaneous symmetry breaking. This conclusion is supported by the observation by Unruh and Weiss that thermal correlation functions computed at a temperature equal to the Unruh temperature are identical to the corresponding correlation functions in a Minkowski (zero-temperature) vacuum. We consider an example of the Nambu-Jona-Lasinio model in a co-accelerating reference frame and show that the uniform acceleration of hot gas of interacting fermions enhances the mass gap generation and increases the critical temperature of the chiral transition. We derive a simple dependence of the critical temperature of a second-order phase transition on acceleration that, as we argue, should be applicable to a broad range of field theories.
\end{abstract}

\maketitle

\paragraph*{\bf Introduction.} 

Acceleration, despite its apparent simplicity, is an intriguing type of non-inertial motion.

Uniform acceleration is associated with the emergence of the Rindler event horizon~\cite{Rindler:1966zz} beyond which the events cannot influence an accelerating particle~\cite{Lee1986}. This observation suggests the existence of a relation between acceleration in flat Minkowski spacetime and the physics of black holes that also possess event horizons, separating causally disconnected regions of spacetime~\cite{Gibbons:1976pt, Gibbons:1976es}.

An intriguing quantum phenomenon associated with acceleration, the Unruh effect, implies that an observer, uniformly accelerated in a zero-temperature Minkowski vacuum, should detect a thermal bath of particles characterized by the Unruh temperature~\cite{Unruh:1976db}:
\begin{align}
    T_U = \frac{|a|}{2\pi}\,.
    \label{eq_T_U}
\end{align}
The Unruh effect may be related to the Hawking process of the black hole evaporation that occurs due to the  production of particle pairs near an event horizon~\cite{Hawking1974, Hawking1975}.

In addition, a uniformly accelerated system provides us with a rare example of a non-inertial physical environment that stays in global thermal equilibrium~\cite{Cercignani:2002, Becattini:2012tc}. The system, however, is not uniform: the local temperature $T$, the local four-velocity $u^\mu$, and the local proper acceleration $a^\mu \equiv u^\nu \partial_\nu u^\mu$ of particles accelerated along the direction $z$ are constrained to be particular functions of coordinates (see, {\it e.g.}, a brief summary in Ref.~\cite{Ambrus:2023smm}):
\begin{align} \label{eq_T_u_a}
 T(t,z) & = \frac{T_{0}}{\sqrt{(1 + a_0 z)^{2} - (a_0 t)^{2}}}\,,\\ 
 u^t(t,z)  & = \frac{T(t,z)}{T_0} (1 + a_0 z), 
\quad
 u^z(t,z) = \frac{T(t,z)}{T_0} a_0 t \,, \nonumber\\
    a^t(t,z) & = \frac{T^2(t,z)}{T_0^2} a_0^2 t,
    \quad 
    a^z(t,z) = \frac{T^2(t,z)}{T_0^2} a_0 (1 + a_0 z), \nonumber
\end{align}
with vanishing transverse components for the four-velocity, $u^x = u^y = 0$, and the four-acceleration, $a^x = a^y = 0$. In Eq.~\eqref{eq_T_u_a}, $T_0 = T(t = z = 0)$ is the reference temperature, and $a_0 = a(t = z = 0)$ is the reference acceleration at the spatial plane $z = 0$ at the time moment $t = 0$. The proper acceleration, $\alpha^{\mu} = a^{\mu}(x)/T(x)$, has a constant magnitude, $\alpha^{\mu} \alpha_\mu = - a_0^2/T^2_0$, along the whole world trajectory of a uniformly accelerating particle. 

Acceleration can also affect the thermodynamic ground state of interacting systems and modify their phase diagrams. In other words, besides the trivial kinematic spatial change of temperature in the inhomogeneous accelerating medium~\eqref{eq_T_u_a}, the acceleration can also play a distinct role in thermodynamics of interacting fields. 

For example, it was suggested in Ref.~\cite{Ohsaku:2004rv} that the acceleration leads to the restoration of the chiral symmetry of interacting fermions described by the Nambu--Jona-Lasinio (NJL) model~\cite{Nambu:1961tp, Nambu:1961fr}. This restoration effect was argued to be a natural consequence of the two-step observation that (i) the acceleration is associated with the Unruh temperature~\eqref{eq_T_U}, while (ii) thermal effects lead to the restoration of a spontaneously broken symmetry~\cite{Dolan:1973qd}. A related restoration of a spontaneously broken symmetry driven by acceleration has also been addressed in other physical scenarios~\cite{Ebert:2006bh, Castorina:2012yg, Takeuchi:2015nga, Dobado:2017xxb, Casado-Turrion:2019gbg, Kou:2024dml}. On the other hand, the symmetry restoration due to acceleration has been questioned in Ref.~\cite{Unruh:1983ac}, which concluded that the acceleration of a vacuum itself cannot cause phase transitions. A very recent and broad critical assessment of the behaviour of the broken symmetry due to acceleration in an interacting $\varphi^4$ field theory can be found in Ref.~\cite{Salluce:2024jlj}.

To address this apparent contradiction, it is essential to distinguish between two physically distinct scenarios: 
\begin{itemize}
    \item[(i)] a Minkowski vacuum (with zero temperature and no particles) perceived by an accelerating observer;
    \item[(ii)] an accelerated physical object as seen by an observer who co-accelerates together with that object. The object appears to be static at the local position in the reference frame of the observer. 
\end{itemize}

In the first scenario, there is no physical object that accelerates. It is a detector that accelerates with respect to an empty Minkowski spacetime.

In the second scenario, we have a real physical matter that accelerates and, therefore, that can physically experience effects caused by the acceleration. Below, we consider the second case by investigating the phase diagram of a thermally equilibrated system of accelerated interacting fermionic particles. 

In an earlier study of Ref.~\cite{Benic:2015qha}, it was shown that the scalar condensate should be enhanced with increasing acceleration, implying that acceleration leads to enhancement rather than to restoration of the symmetry breaking in an accelerating medium. In our article, we support the conclusion of Ref.~\cite{Benic:2015qha} by considering the Nambu--Jona-Lasinio model of interacting fermions~\cite{Nambu:1961tp, Nambu:1961fr}. This model is phenomenologically relevant to heavy-ion collisions where the quark-gluon system is expected to be subjected to strong accelerations~\cite{Kharzeev:2005iz}.

In the following, we consider a positive acceleration $a_0 > 0$ and redefine, for notational simplicity, the reference quantities in Eq.~\eqref{eq_T_u_a} as $a_0 \to a$ and $T_0 \to T$. We work at a fixed temperature $T$ and acceleration $a$ that correspond to the $t = z = 0$ plane. The obtained results can be further extended to the whole Rindler wedge by promoting the constant parameters $T$ and $a$ to the coordinate-dependent quantities $T \to T(t,z)$ and $a \to a(t,z)$ following Eq.~\eqref{eq_T_u_a} which assures the global thermal equilibrium of the system~\cite{Becattini:2019poj, Becattini:2020qol, Selch:2023pap}. 

\vskip 3mm
\paragraph*{\bf Acceleration as refrigeration: free scalar theory.} The essence of the discussed acceleration-assisted cooling effect can be demonstrated by an example of a simple bubble diagram for a free massless scalar field.  A thermal contribution to the quadratic fluctuations of the massless field in the theory without acceleration reads as follows:
\begin{align}
\delta \langle \phi^2 \rangle_T = \int \frac{d^3 p}{(2\pi)^3} \frac{f^{(B)}_{\boldsymbol{p}}} {\epsilon_{\boldsymbol{p}}} 
  & = \frac{T^2}{12}\,,
  \label{eq_thermal_B}
\end{align}
where $f^{(B)}_{\boldsymbol{p}} = 1/(e^{\epsilon_{\boldsymbol{p}}/T} - 1)$ is the Bose-Einstein distribution function for a massless boson with the energy $\epsilon_{\boldsymbol{p}} = |\boldsymbol{p}|$. Equation~\eqref{eq_thermal_B} represents a thermal part of a one-loop correction to the boson propagator, which, in turn, generates a thermal correction to an effective mass of the boson. As a result, an increasing temperature leads to the restoration of spontaneously broken continuous symmetries~\cite{Dolan:1973qd}.

In an accelerating medium, the thermal quadratic fluctuations of the scalar field decrease~\cite{Moretti:1997qn, Becattini:2020qol, Diakonov:2023hzg, Ambrus:2023smm}:
\begin{align}
	\delta \langle \phi^2 \rangle_{T,a} = \frac{1}{12} \Bigl(T^2 - \frac{a^2}{4 \pi^2}\Bigr)\,.
 \label{eq_phi2_T_a}
\end{align}
This relation is re-derived in the Appendix. 

Equation~\eqref{eq_phi2_T_a} demonstrates that the acceleration $a$ diminishes the thermal mass correction and, consequently, reduces the effect of temperature $T$ on thermal fluctuations of the scalar field. Exactly the same conclusion has been reached in Ref.~\cite{Benic:2015qha}, where an acceleration-generated negative correction to quadratic fluctuations of the scalar field has been found. It was argued that the scalar condensate ---if the theory exhibits a spontaneous symmetry breaking--- should increase as the acceleration gets larger~\cite{Benic:2015qha}.

In our article, we interpret the effect of acceleration in Eq.~\eqref{eq_phi2_T_a} as a cooling effect because the influence of acceleration of a thermal medium of scalar particles that accelerate in the global thermal equilibrium can be interpreted as a temperature shift:
\begin{align}
	T^2 \to T^2 - \frac{a^2}{4 \pi^2}\,.
 \label{eq_T_shift}
\end{align}
In other words, the fluctuations of uniformly accelerating fields are governed by the shifted temperature~\eqref{eq_T_shift} rather than by the local temperature~$T$. 

The modification of temperature~\eqref{eq_T_shift} should emerge not only in a free theory, but it should also appear in loop calculations for interacting fields. For example, one finds the temperature shift~\eqref{eq_T_shift} also in perturbative computations in the scalar field theory with $\lambda \phi^4$ self-interaction in the Rindler spacetime~\cite{Diakonov:2023hzg}. Therefore, formula~\eqref{eq_T_shift} can have a universal nature.

In theories that exhibit a spontaneous symmetry breaking at low temperature, the effect of thermal fluctuations is known to lead to the restoration of the broken symmetry~\cite{Dolan:1973qd}. Due to the acceleration-assisted cooling effect~\eqref{eq_T_shift}, it is natural, therefore, to claim that the acceleration acts oppositely to the effect of thermal fluctuations by enhancing the spontaneous symmetry breaking. Below, we demonstrate this phenomenon using the NJL model~\cite{Nambu:1961tp, Nambu:1961fr} as an example.

\vskip 3mm
\paragraph*{\bf The Unruh-Weiss observation.} The effect of acceleration on quadratic fluctuations of the scalar field~\eqref{eq_phi2_T_a} is tightly connected with the notion of the Unruh temperature~\eqref{eq_T_U}, which is experienced by an observer undergoing uniform acceleration through a vacuum. When the uniform acceleration reaches the critical value $a_c = a_T = 2 \pi T$, the Unruh temperature~\eqref{eq_T_U} becomes equal to the temperature of the medium, $T_U = T$. As a consequence, the thermal correction~\eqref{eq_phi2_T_a} to the quadratic fluctuations of the scale field ${\cal O} = \phi^2$ vanishes:
\begin{align}
	\delta \langle {\cal O} \rangle_{T,a}{\Bigl|}_{T = T_U \equiv a/2 \pi} = 0\,.
    \label{eq_Unruh_Weiss}
\end{align}
This result is in line with the Unruh-Weiss observation~\cite{Unruh:1983ac}, valid for both free and interacting Lorentz-invariant field theories, which states that the ``accelerated observer in thermal equilibrium at $T = a/2\pi$ measures the same Green's functions as the Minkowski observer in his vacuum state.'' Relation~\eqref{eq_Unruh_Weiss} has a rather generic nature because it applies to other quantities, including the energy-momentum tensor in various models (${\cal O} = T^{\mu\nu}$), all components of which vanish at the Unruh temperature $T = T_U$~\cite{Moretti:1997qn, Becattini:2020qol, Palermo:2021hlf, Diakonov:2023hzg}. The Unruh-Weiss statement is also consistent with our interpretation of the cooling effect of the acceleration~\eqref{eq_T_shift}.

If one continues to accelerate the medium with a fixed temperature $T$ above the threshold value $a_T = 2 \pi T$, then the quadratic fluctuations~\eqref{eq_phi2_T_a} become negative. While, very formally, this observation goes in line with the mentioned symmetry-breaking property of acceleration, it leads to a difficulty of its physical interpretation. Interestingly, this observation may be related to an acceleration-induced phase transition at temperature $T = T_U$ suggested recently in Ref.~\cite{Prokhorov:2023dfg}. In our article, we do not attempt to touch this subtle matter, and we restrict ourselves to temperatures in the range $T \geqslant T_U$. 

\vskip 3mm
\paragraph*{\bf The Nambu-Jona-Lasinio (NJL) model\!\!} describes interacting relativistic fermions:
\begin{align}
\mathcal{L}_{{\rm NJL}} = & \, \bar{\psi}(i\gamma^\mu \partial_\mu - m_0)\psi + G\bigl[(\bar{\psi}\psi)^2 + (\bar{\psi}i\gamma_5\psi)^2\bigr]\,,
\label{eq_L_NJL}
\end{align}
where $\psi$ is the fermion field with the bare mass $m_0$ and $G$ is a dimensionful coupling constant that controls the fermionic four-point interactions. 

In the regime when the chiral symmetry is spontaneously broken, the fermionic interaction generates a nonzero constituent fermion mass $M$ even if the bare fermion mass is vanishing. Therefore, for simplicity, we consider the chiral limit of the NJL model~\eqref{eq_L_NJL} by setting the bare mass to zero, $m_0 = 0$. The dynamical mass $M$, which is induced by the dynamically generated fermion condensate $\langle \bar{\psi}\psi \rangle$, can be computed in a standard way via a gap equation, which, in turn, is derived by minimizing the thermodynamic potential with respect to the dynamical fermion mass $M$. 

\vskip 3mm
\paragraph*{\bf The mass gap equation in the vacuum}\!\!\!\!\! can conveniently be computed in the mean-field approximation by linearizing the four-fermion interaction:
\begin{align}
\mathcal{L}_{{\rm MF}} = \bar{\psi}(i\gamma^\mu \partial_\mu - M)\psi - \frac{M^2}{4G}\,.
\end{align}
The parameter $M$ is treated as an auxiliary scalar field, which represents the dynamical fermion mass. The second four-fermion term in the squared brackets in Eq.~\eqref{eq_L_NJL} is traditionally neglected since the associated condensate is vanishing in the absence of chirality-related backgrounds.

In the mean-field approximation, the effective thermodynamic potential $\Omega(M)$ in the NJL model is obtained by integrating out the fermion fields:
\begin{align}
\frac{\Omega(M)}{V_4} & = \frac{M^2}{4G} - i \, {\rm Tr} \, \ln \left[ i\gamma^\mu \partial_\mu - M \right] 
\nonumber \\
& = \frac{M^2}{4G} - 2 \int \frac{d^4 p}{(2\pi)^4} \ln(p^2 - M^2 + i\epsilon)\,,
\label{eq_Omega_1}
\end{align}
where the second term arises from the fermion loop evaluated in the momentum space, $V_4$ is the spacetime volume, and $p^2 = p_0^2 - {\boldsymbol{p}}^2$. After performing the Wick rotation to Euclidean space $p_0 \rightarrow ip_4$, and denoting $p_E^2 = p_4^2 + \boldsymbol{p}^2$, the thermodynamic potential~\eqref{eq_Omega_1} at zero temperature takes the following form:
\begin{align}
\frac{\Omega(M)}{V_4} = \frac{M^2}{4G} - 2 \int \frac{d^4 p_E}{(2\pi)^4} \ln(p_E^2 + M^2)\,.
\label{eq_Omega_2}
\end{align} 

The minimization of the thermodynamic potential~\eqref{eq_Omega_2} with respect to the variation of the mass parameter, $\partial \Omega(M)/\partial M = 0$, gives us the gap equation:
\begin{align}
    M = -2G \langle \bar{\psi}\psi \rangle_0\,, 
    \label{eq_gap_1}
\end{align}
which relates the dynamically generated mass $M$ to the fermion condensate:
\begin{align}
\langle \bar{\psi}\psi \rangle_0 = -2 \int^{\Lambda} \frac{d^3 p}{(2\pi)^3} \frac{M}{E_{\boldsymbol{p}}}\,.
\label{eq_condensate}
\end{align}
Here $E_{\boldsymbol{p}} = \sqrt{\boldsymbol{p}^2 + M^2}$ is the fermion energy. The subscript ``0'' in Eqs.~\eqref{eq_gap_1} and \eqref{eq_condensate} indicates that here, the condensate is evaluated at zero temperature $T = 0$ (and, obviously, in the absence of acceleration $a = 0$). 

The integration in Eq.~\eqref{eq_condensate} is divergent in the ultraviolet limit, $p \to \infty$, and, therefore, the integral requires an appropriate regularization. A commonly used regularization scheme in the NJL model is the three-momentum cutoff $\Lambda$, which imposes a limiting value on the spatial momentum integral. The cutoff $\Lambda$ plays a role of a physical parameter of the model, which has to be fixed from phenomenological considerations~\cite{Klevansky:1992qe}. Importantly, the cutoff $\Lambda$ also sets the mass scale in the system.

Thus, the gap equation~\eqref{eq_gap_1} becomes a self-consistent equation for the dynamical mass $M$:
\begin{align}
1 = 4 G \int^{\Lambda} \frac{d^3 p}{(2\pi)^3} \frac{1}{E_{\boldsymbol{p}}}\,.
\label{eq_gap_2}
\end{align}
This integral can be taken explicitly, giving us
\begin{subequations}
\begin{align}
1 & = G \Lambda^2 I\biggl(\frac{M}{\Lambda}\biggr)\,,
\label{eq_gap_3}\\
I(x) & = \frac{1}{\pi^2} \Bigl[\sqrt{1 + x^2} - x^2 {\rm arcsinh}(1/x)\Bigr] \nonumber\\
& = \frac{1}{\pi^2} \Bigl(1 - x^2 \ln \frac{x \sqrt{e}}{2}\Bigr) + O(x^4)\,.
\label{eq_f}
\end{align}
\label{eq_mass_gap_T0}
\end{subequations}
\!\!At the last relation in Eq.~\eqref{eq_f}, we presented the integral $I(x)$ in terms of a series over a small argument~$x$. 

The vacuum mass gap equation~\eqref{eq_mass_gap_T0} can be used to determine the critical vacuum value of the coupling $G$:
\begin{align}
    G^{\rm vac}_c = \frac{\pi^2}{\Lambda^2}\,. 
\label{eq_Gc_vac}
\end{align}
Thus, we arrive at a well-known conclusion that at zero temperature, the NJL model~\eqref{eq_L_NJL} has a two-phase structure. At strong coupling ($G > G_c^{{\rm vac}}$), the system resides in the symmetry-broken phase characterized by a non-zero dynamical mass $M \neq 0$ and non-vanishing chiral condensate, $\langle \bar\psi\psi\rangle \neq 0$. At weak coupling ($G \leqslant G_c^{{\rm vac}}$), a symmetry restoration takes place, implying $M=0$ and $\langle \bar\psi\psi\rangle = 0$. The phase transition is of a second order, indicating that in the vicinity of the critical coupling, $G \simeq G_c^{{\rm vac}}$, the mass gap $M$ is small, $M \ll \Lambda$.

\vskip 3mm
\paragraph*{\bf The mass gap equation at finite temperature.}
At finite temperature $T$, the mass gap equation~\eqref{eq_mass_gap_T0} acquires the contribution from thermal fluctuations,
\begin{align}
1 = 4 G \int^{\Lambda} \frac{d^3 p}{(2\pi)^3} \frac{1}{E_{\boldsymbol{p}}} \Bigl(1 - 2 f^{(F)}_{\boldsymbol{p}}\Bigr)\,,
\label{eq_gap_T_1}
\end{align}
where $f^{(F)}_{\boldsymbol{p}} = 1/(e^{E_{\boldsymbol{p}}/T}+1)$ is the Fermi-Dirac distribution. The integral over the thermal part is converging, and therefore, the cutoff $\Lambda$ in the thermal integral can be lifted off, $\Lambda \to \infty$. In a low-mass regime, $M \ll T$, the finite-temperature correction to the condensate, ${\langle {\bar \psi} \psi \rangle}_T = {\langle {\bar \psi} \psi \rangle}_{0} + \delta {\langle {\bar \psi} \psi \rangle}_T$, reads as follows~\cite{Laine:2016hma}:
\begin{align}
    	\delta {\langle {\bar \psi} \psi \rangle}_T = 4 M \int \frac{d^3 p}{(2\pi)^3} \frac{f^{(F)}_{\boldsymbol{p}}}{E_{\boldsymbol{p}}} 
  & = \frac{M T^2}{6} \,,
  \label{eq_condensate_T}
\end{align}
where the subleading $O(M^2 T)$ terms were omitted.

We will be working in the same low-mass regime where the dynamically generated fermion mass is much smaller than temperature. In this regime, corresponding to a vicinity of the second-order transition region, 
the mass-gap equation~\eqref{eq_gap_1} acquires the following simple form:
\begin{align}
	1 & = G \Lambda^2 \Biggl[I\biggl(\frac{M}{\Lambda}\biggr) - \frac{T^2}{3 \Lambda^2}  \Biggr]\,.
\label{eq_gap_T}
\end{align}
This equation is analogous to the vacuum mass gap equation~\eqref{eq_gap_1}, in which the fermionic condensate at the right-hand side has been substituted by the sum of the original vacuum term~\eqref{eq_condensate} and the thermal contribution~\eqref{eq_condensate_T}:
\begin{align}
    M = -2G \bigl( \langle \bar{\psi}\psi \rangle_0 + \delta \langle \bar{\psi}\psi \rangle_T \bigr)\,.
    \label{eq_gap_T_2}	
\end{align}
Since the function $f(x)$ decreases with $x$ for a small argument~\eqref{eq_f}, the mass gap equation~\eqref{eq_gap_T} implies that the thermal fluctuations reduce the dynamical mass and, consequently, destroy the chiral condensate. 

A solution of the mass gap equation~\eqref{eq_gap_T} gives us another well-known result, now for the critical temperature of the chiral restoration transition:
\begin{align}
    T_{c0} = \sqrt{3 \biggl(\frac{1}{G_c^{\rm vac}} - \frac{1}{G} \biggr)}\,.
    \label{eq_T0_c}
\end{align}
Thus, in order to support the chiral symmetry breaking at finite temperature, the four-fermion coupling $G$ should be stronger than the critical vacuum value $G_c^{\rm vac}$ given in Eq.~\eqref{eq_Gc_vac}. The additional subscript ``0'' in $T_{c0}$ in indicates that the critical temperature $T_{c0}$ corresponds to the phase transition in the absence of acceleration, $a = 0$. Now, let us consider how acceleration affects the critical temperature of the chiral symmetry restoration.

\vskip 3mm
\paragraph*{\bf The mass gap equation at finite temperature and acceleration.}
The mass gap equation in an accelerating medium, $M = -2G (\langle \bar{\psi}\psi \rangle_0 + \delta \langle \bar{\psi}\psi \rangle_{T, a})$, includes the fermionic condensate given by the sum of the original vacuum term~\eqref{eq_condensate} and the contribution of the accelerating thermal medium $\delta {\langle {\bar \psi} \psi \rangle}_{T,a}$. The latter can be evaluated explicitly: 
\begin{align}
	\delta {\langle {\bar \psi} \psi \rangle}_{T,a} = \frac{M}{6} \Bigl(T^2 - \frac{a^2}{4 \pi^2}\Bigr)\,.
 \label{eq_condensate_T_a} 
\end{align}
The details of the derivation are given in the Appendix.

The condensate~\eqref{eq_condensate_T_a}  follows the general rule of the cooling effect imposed by the acceleration of the thermal medium~\eqref{eq_T_shift}. According to our approach, Eq.~\eqref{eq_condensate_T_a}  is valid in the limit of a small dynamical mass, $M \ll T$, which applies to a vicinity of the second-order phase transition. This approximation is enough for our purposes to identify the effect of acceleration on the critical temperature.

Thus, the mass gap equation of an accelerating gas of hot interacting fermions is given by:
\begin{align}
	1 & = G \Lambda^2 \Biggl[I\biggl(\frac{M}{\Lambda}\biggr) - \frac{1}{3 \Lambda^2}\biggl(T^2 - \frac{a^2}{4 \pi^2}\biggr) \Biggr]\,,
\label{eq_gap_T_3}
\end{align}
where, again, the vacuum part is represented by Eq.~\eqref{eq_f}. At the point of the second-order phase transition, the mass gap vanishes, $M =0$. Then, using $I(0) = 1/\pi^2$ in Eq.~\eqref{eq_gap_T}, we get the following dependence of the critical temperature on the acceleration $a$:
\begin{align}
	T_c(a) = \sqrt{T_{c0}^2 + \frac{a^2}{4 \pi^2}}\,,
\label{eq_Tc_a_0}
\end{align}
where $T_{c0}$ is the critical temperature of the non-accelerating fermionic matter~\eqref{eq_T0_c}. One can rewrite this equation in more concise notations:
\begin{align}
	T_c(a) = \sqrt{T_{c0}^2 + T_U^2(a)}\,,
 \label{eq_Tc_TU}
\end{align}
where $T_U = T_U(a)$ is the Unruh temperature~\eqref{eq_T_U}. Notice that the dependence of the critical temperature $T_c$ on acceleration $a$, Eq.~\eqref{eq_Tc_a_0} or Eq.~\eqref{eq_Tc_TU}, has a simple universal form that does not depend on the structure of the underlying theory (in our case, it is the NJL model). Following the Unruh-Weiss arguments~\cite{Unruh:1983ac}, we expect that the critical law~\eqref{eq_Tc_TU} works in a wide class of Lorentz-invariant field theories, at least for phase transitions of the second order (for which the mass gap $M$ vanishes at $T = T_{c0}$). 

\begin{figure}
    \includegraphics[width = 1.0\linewidth]{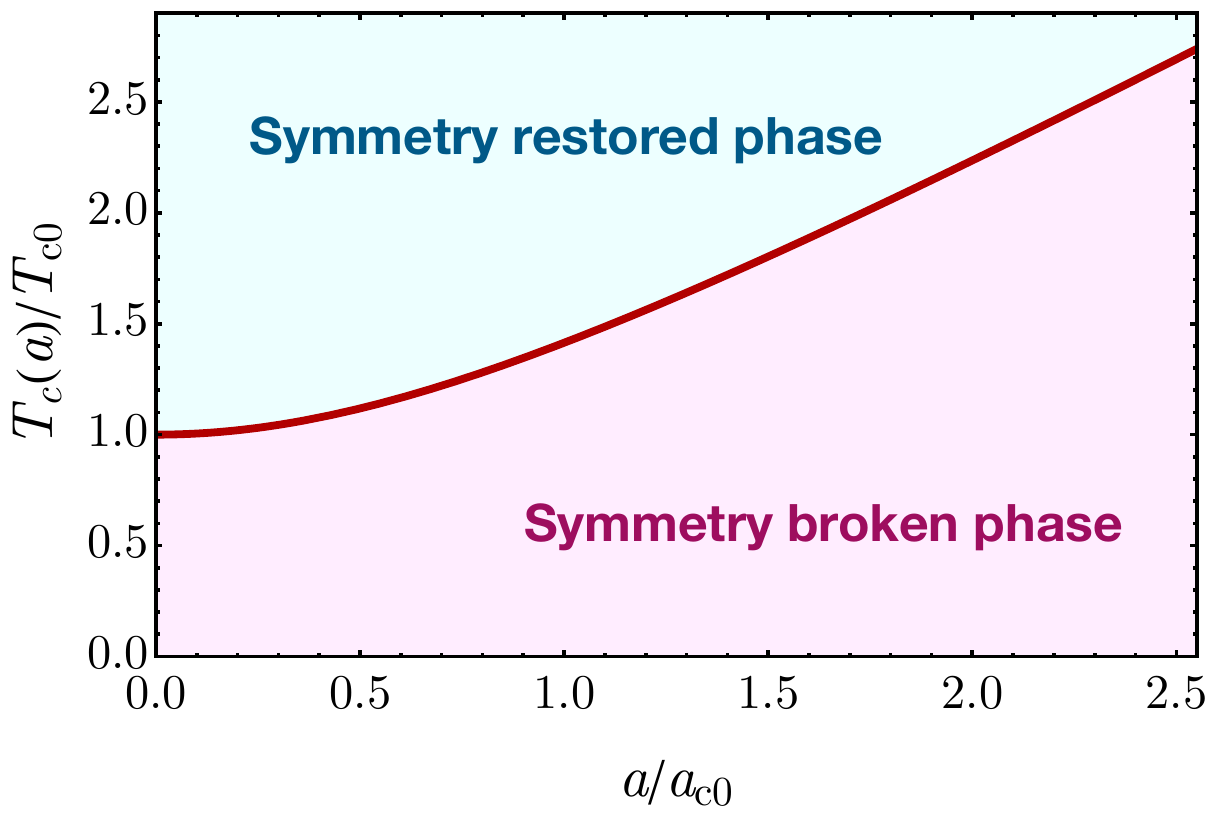}
    \caption{Dependence~\eqref{eq_Tc_a} of the critical temperature $T_c(a)$ of symmetry restoration on acceleration $a$. The critical temperature is given in units of the critical temperature $T_{c0}$ at a vanishing acceleration, $a=0$, Eq.~\eqref{eq_T0_c}. The acceleration $a$ is given in units of the characteristic acceleration scale, $a_{c0}$, Eq.~\eqref{eq_a_c0}, which also depends on $T_{c0}$. }
    \label{fig}
\end{figure}

Using Eq.~\eqref{eq_a_c0}, the effect of acceleration on the critical temperature~\eqref{eq_Tc_a_0} can be put in a more transparent way:
\begin{align}
	\frac{T_c(a)}{T_{c0}} = \sqrt{1 + \biggl(\frac{a}{a_{c0}}\biggr)^2}\,,
\label{eq_Tc_a}
\end{align} 
where we introduced the characteristic acceleration scale 
\begin{align}
	a_{c0} = 2 \pi T_{c0}\,,
    \label{eq_a_c0}
\end{align}
corresponding to the critical temperature in the absence of acceleration~$T_{c0}$. The critical temperature~\eqref{eq_Tc_a} is shown in Fig.~\ref{fig}. The acceleration increases the critical temperature of the phase transition, thus providing an effect similar to cooling. 

Notice that the critical transition temperature~\eqref{eq_Tc_a} can never reach the Unruh temperature~\eqref{eq_T_U} due to the strict inequality $T_{c}(a) > T_U(a)$. Thus, the Unruh-Weiss argument~\cite{Unruh:1983ac} does not apply to the vicinity of the acceleration-modified critical temperature~\eqref{eq_Tc_a}, and does not spoil our arguments.

\vskip 1mm
\paragraph*{\bf A remark on numerical results on accelerating gluon gas in lattice Yang-Mills theory.} 
Recently, a numerical Monte Carlo calculation of the phase diagram of accelerating gluon gas in finite-temperature lattice Yang-Mills theory has been performed~\cite{Chernodub:2024wis}. The position of the transition temperature was found to be unaffected by acceleration within a 1\% accuracy. The absence of the noticeable numerical effect on the deconfinement temperature persisted for the accelerations up to the maximal studied value of $a^{\rm max}_{\rm lat} \simeq 27$\,MeV.

This numerical result might appear to be puzzling given that the studied acceleration has a substantial magnitude, about 10\% of the critical deconfinement temperature of SU(3) Yang-Mills theory, $a^{\rm max}_{\rm lat}/T^{SU(3)}_{c0} \simeq 0.1$. The latter quantity, at a vanishing acceleration, is estimated to be $T^{SU(3)}_{c0} = 0.629(3) \sqrt{\sigma_0} \simeq 305$\,MeV~\cite{Boyd:1996bx}, if one takes the value of the zero-temperature string tension as $\sqrt{\sigma_0} = 486(6)\mathrm{\,MeV}$~\cite{Athenodorou:2020ani}. 

However, the puzzling numerical result of Ref.~\cite{Chernodub:2024wis} is consistent with our main formula~\eqref{eq_Tc_a}. The apparent insensitivity of the deconfinement temperature on acceleration is a result of a very shallow dependence $T_c = T_c(a)$ on $a$ at weak accelerations, $|a| \ll T_{c0}$. For the maximal value of acceleration used in the numerical simulations, the theoretical shift of the critical temperature~\eqref{eq_Tc_a} is, in fact, tiny:
\begin{align}
		\frac{T_c(a^{\rm max}_{\rm lat})}{T^{SU(3)}_{c0}} = \sqrt{1 + \biggl(\frac{a^{\rm max}_{\rm lat}}{2 \pi T^{SU(3)}_{c0}} \biggr)^2} = 1.0000993\dots \,.
\label{eq_Tc_a_SU3}
\end{align}
Numerically, the modification of the deconfining transition temperature at the studied values of acceleration should theoretically be less than $0.01\%$ which is much smaller than the numerical accuracy (about~1\%) of the numerical Monte Carlo calculation performed in Ref.~\cite{Chernodub:2024wis}. Thus, our formula~\eqref{eq_Tc_a} is consistent with the results of the lattice simulations.

\vskip 3mm
\paragraph*{\bf Conclusions.} We argued that the uniform acceleration of hot matter produces a cooling effect on that matter. Working in the NJL model, we derived the dependence of the critical temperature on acceleration, given by Eqs.~\eqref{eq_Tc_a} and \eqref{eq_a_c0}. We suggest that this temperature behaviour has a universal character valid in a wide class of field theories.

\vskip 3mm
\paragraph*{\bf Acknowledgments.} The author is grateful to Victor Ambru\cb{s} for useful comments and initial collaboration on this material, and to Francesco Becattini, Antonino Flachi, and Kenji Fukushima, for useful discussions and critical remarks. This work was funded by the EU’s NextGenerationEU instrument through the National Recovery and Resilience Plan of Romania - Pillar III-C9-I8, managed by the Ministry of Research, Innovation and Digitization, within the project entitled ``Facets of Rotating Quark-Gluon Plasma'' (FORQ), contract no.~760079/23.05.2023 code CF 103/15.11.2022. 

\appendix
\section{~\\ Acceleration effects for scalar and fermionic fields}

Evaluation of the acceleration effects on the local expectation values of the field operators requires certain care. Therefore, it is necessary to discuss them in more detail by highlighting some important steps in the derivation. In this Appendix, we give a brief description of methods applied to accelerating bosonic (fermionic) fields and related quadratic (bilinear) field observables in thermodynamic ground state of field theories. We also explicitly derive Eqs.~\eqref{eq_thermal_B}, \eqref{eq_phi2_T_a}, \eqref{eq_condensate_T}, and \eqref{eq_condensate_T_a}. For a more complete and exhaustive description of field theories under acceleration, we refer an interested reader, for example, to Refs.~\cite{Prokhorov:2019hif, Prokhorov:2019cik, Zakharov:2020ked, Becattini:2020qol, Palermo:2021hlf, Ambrus:2023smm}.

\vskip 1mm
\paragraph*{\bf Scalar fields.}

Let us start with the simplest case of a massless scalar field theory described by the Lagrangian:
\begin{align}
	{\cal L}_\phi = \frac{1}{2} \int d^4 x \, \partial_\mu \phi \partial^\mu \phi\,.
    \label{eq_L_phi}
\end{align}
Since we consider thermal equilibrium, it is convenient to perform the Wick rotation to imaginary time, $t \to \tau = i t$, and formulate theory~\eqref{eq_L_phi} at the Euclidean manifold. 

\vskip 1mm
\paragraph*{---Zero temperature $(T=0)$, no acceleration $(a = 0)$.} The expectation value of the quadratic fluctuations of the scalar field is:
\begin{align}
	\langle \phi^2 \rangle_0 \equiv \langle \phi^2(0) \rangle_0 = \lim_{x \to 0} \langle \phi(0) \phi(x) \rangle_0\,,
    \label{eq_phi_2}
\end{align}
where $x = ({\boldsymbol{x}},\tau)$ is the Euclidean coordinate. The subscript ``$0$'' in Eq.~\eqref{eq_phi_2} means that the expectation value is taken at zero temperature. 

For our purposes, it is convenient to perform all calculations in the coordinate space. At zero temperature, the Euclidean correlation function is given by the Green function of a Laplacian operator in the ${\mathbb R}^4$ space,
\begin{align}
	G_0(x) \equiv \langle \phi(0) \phi(x) \rangle = \frac{1}{4 \pi^2} \frac{1}{|x|^2}\,,
    \label{eq_G_vac}
\end{align}
where $|x|^2 = {\boldsymbol{x}}^2 + \tau^2$. The quadratic divergence of the Green function~\eqref{eq_G_vac} in the local limit, $x \to 0$, implies that the zero-temperature correlation function~\eqref{eq_phi_2} diverges quadratically as well. 

\vskip 1mm
\paragraph*{---Finite temperature $(T \neq 0)$, no acceleration $(a = 0)$.} The finite-temperature correction to quadratic fluctuations of the scalar field $\phi$ is defined as a difference in fluctuations at zero and finite temperatures: 
\begin{align}
	\delta \langle \phi^2 \rangle_T = \lim_{x \to 0} 
    \bigl(\langle \phi(0) \phi(x) \rangle_T - \langle \phi(0) \phi(x) \rangle_0 \bigr)\,.
    \label{eq_phi_2_delta}
\end{align}
To calculate this quantity, let us recall that the thermal expectation values of a field theory in thermal equilibrium at temperature $T$ can be evaluated in the Euclidean spacetime with the imaginary time coordinate compactified to a circle of the length $1/T$, as follows from the Kubo-Martin-Schwinger (KMS) condition for bosons. At finite temperature, all the points
\begin{align}
    & x_{(n)} = \bigl( {\boldsymbol{x}}, \tilde{\tau}_{(n)} \bigr) \quad {\rm with} 
    \quad 
    \tilde{\tau}_{(n)} = \tau + \frac{n}{T},
    \quad
    n \in {\mathbb Z}\,,
    \label{eq_x_n_T}\\
    & \hskip 35mm ({\rm for}\ \ T \neq 0\ \ {\rm and} \ \  a = 0)\,, \nonumber
\end{align}
are identified, resulting in the following relation for the bosonic Euclidean Green function:
\begin{align}
	G_T({\boldsymbol{x}},\tilde{\tau}_{(n)}) = G_T({\boldsymbol{x}},\tau_{(n')})\,, \qquad \forall n,n' \in {\mathbb Z}\,,
\end{align}
where the integers $n$ and $n'$ label replicas along the imagiary time direction. We also took into account the periodicity of the bosonic fields under the imaginary time translations, $\tau \to \tau + 1/T$, which follows from their Bose-Einstein statistics.

The thermal analogue of the Green function~\eqref{eq_G_vac} can therefore be obtained with the method of reflections with respect to the imaginary time coordinate~$\tau$:
\begin{align}
	G_T({\boldsymbol{x}},\tau) = & \sum_{n \in {\mathbb Z}} G_0({\boldsymbol{x}},\tau + n/T)\,,
    \label{eq_G_T}
\end{align}
where the sum goes over the mirrored replicas along the time direction. A straightforward evaluation gives us:
\begin{align}
    G_T({\boldsymbol{x}},\tau) = & \frac{1}{4 \pi^2} \sum_{n \in {\mathbb Z}} \frac{1}{|{\boldsymbol{x}}|^2 + (\tau + n/T)}
    \nonumber \\
    = & \frac{T}{4 \pi |{\boldsymbol{x}}|} \frac{\sinh (2 \pi T |{\boldsymbol{x}}|)}{\cos(2 \pi T \tau) - \cosh (2 \pi T |{\boldsymbol{x}}|)}\,.
    \label{eq_G_T_explicit}
\end{align}

Using Eqs.~\eqref{eq_G_vac} and \eqref{eq_G_T_explicit} one arrives at the conclusion that the thermal contribution to the expectation value~\eqref{eq_phi_2_delta} is a finite quantity given by Eq.~\eqref{eq_thermal_B} of the main text:
\begin{align}
	\delta \langle \phi^2 \rangle_T = \lim_{{}^{\ \tau \to 0}_{|{\boldsymbol{x}}| \to 0}} \Bigl( G_T({\boldsymbol{x}},\tau) - G_0({\boldsymbol{x}},\tau) \Bigr) = \frac{T^2}{12}\,.
    \label{eq_phi_2_T}
\end{align}

\vskip 1mm
\paragraph*{---Finite temperature $(T {\neq} 0)$ and acceleration $(a {\neq} 0)$.} 
The KMS boundary condition for a bosonic system accelerating uniformly along the direction $z$ leads to the following identification of the coordinates~\cite{Ambrus:2023smm}:
\begin{align}
    x_{(n)} & = ({\boldsymbol{x}}_\perp, z_{(n)},\tau_{(n)}) \quad {\rm with} \nonumber \\
 \tau_{(n)} & = \tau \cos(n \alpha) - \frac{1}{a} (1 + a z) \sin(n \alpha), 
 \label{eq_x_n_T_a}\\
    z_{(n)} & = \tau \sin(n \alpha) + \frac{1}{a} (1 + a z) \cos(n \alpha) - \frac{1}{a}, \nonumber\\
    & \hskip 30mm ({\rm for}\ \ T \neq 0\ \ {\rm and} \ \  a \neq 0)\,, \nonumber
\end{align}
where
\begin{align}
	\alpha = \frac{a}{T}\,,
    \label{eq_alpha}
\end{align}
is the dimensionless thermal acceleration. The coordinate ${\boldsymbol{x}}_\perp$ corresponds to the spatial transverse plane, which is normal to the acceleration direction. Equation~\eqref{eq_x_n_T_a} is a $a \neq 0$ analogue of Eq.~\eqref{eq_x_n_T}. Without restriction of generality, it is convenient to consider positive acceleration values, $a = |a| > 0$. 

The KMS identification of the points~\eqref{eq_x_n_T_a} of the uniformly accelerating medium corresponds to an infinite set of rotations in the $(z, \tau)$ plane about the point $(z,\tau)_R = (- 1/a, 0)$. Amusingly, the rotations are quantized in terms of an elementary angle equal to the thermal acceleration $\alpha$ given in Eq.~\eqref{eq_alpha}. This unusual property can also be interpreted as if the field theory is formulated in the background of a cosmic string defect~\cite{Dowker1987, Linet1995}  with the angle deficit $\Delta \varphi = 2 \pi - \alpha$ related to the thermal acceleration $\alpha$~\cite{Moretti:1997qn, Prokhorov:2019hif, Prokhorov:2019yft, Zakharov:2020ked}.

The point $(z,\tau)_R = (- 1/a, 0)$ of the Euclidean spacetime corresponds to the Euclidean Rindler horizon~\cite{Ambrus:2023smm}. We recall that in Minkowski spacetime, a uniformly accelerating particle travels along a hyperbolic trajectory with the entire worldline confined within the right Rindler wedge, $z > |t| - 1/a_0$. The boundary of the Rindler wedge corresponds to the Rindler horizon, $z_{\rm R}(t) = |t| - 1/a_0$, which defines the Rindler event horizon where all quantities~\eqref{eq_T_u_a} diverge. After the Wick transformation, the boundary of the Rindler wedge of the Minkowski space shrinks to the single point $(z,\tau)_R$ in the Euclidean space. In the static, no-acceleration limit, $a \to 0$, the Euclidean Rindler horizon moves to a spatial infinity, and the KMS identification~\eqref{eq_x_n_T_a} reduces to the standard Matsubara form~\eqref{eq_x_n_T}.

The Green function corresponding to the accelerating thermal bosonic gas~\cite{Birrell1984, Ambrus2021},
\begin{align}
 G_{T,a}({\boldsymbol{x}}_\perp, z, \tau) & = \sum_{n \in {\mathbb Z}} G_0({\boldsymbol{x}}_\perp, z_{(n)}, \tau_{(n)})\,,
\end{align}
is a formally divergent quantity. Performing the steps identical to the non-accelerating thermal case, we arrive at the following result for the quadratic fluctuations of the scalar field in the accelerating bosonic gas:
\begin{align}
	\delta \langle \phi^2 \rangle_{T,a}  = \frac{T^2}{8 \pi^2} S_2(i \alpha)\,,
    \label{eq_delta_phi_2_T_a_sum}
\end{align}
where, in notations of Ref.~\cite{Becattini:2020qol}:
\begin{align}
	S_2 (\phi) = \sum_{n = 1}^\infty 
    \frac{\phi^2}{\sinh^2 (\phi n/2)}\,.
\end{align}
This sum can be evaluated using an analytic distillation procedure. Mathematically, this procedure regularizes the sum by selecting an analytical non-divergent contribution~\cite{Becattini:2020qol}. Physically, the analytic distillation calculates the thermal expectation value of the field operators in co-accelerating (and also, co-rotating) frames. For acceleration, it gives an expectation value that satisfies the Unruh-Weiss normalization point at $T = T_U$~\cite{Unruh:1983ac}, Eq.~\eqref{eq_Unruh_Weiss}.

The analytically distilled value of the $S_2 (\phi)$ polynomial gives us (see Appendix~C of Ref.~\cite{Becattini:2020qol}):
\begin{align}
	S_2(\phi) = \frac{1}{6} \Bigl(\phi^2 + (2\pi)^2 \Bigr)\,,
\end{align}
so that we arrive to Eq.~\eqref{eq_phi2_T_a} of the main text:
\begin{align}
	\delta \langle \phi^2 \rangle_{T,a}  = \frac{1}{12} \biggl[ T^2 - \Bigl(\frac{a}{2\pi}\Bigr)^2 \biggr]\,.
    \label{eq_delta_phi_2_T_a_2}
\end{align}
At zero acceleration, $a = 0$, the above formula gives us the known result~\eqref{eq_phi_2_T}. Notice that Eq.~\eqref{eq_delta_phi_2_T_a_2} can also be obtained using other methods~\cite{Moretti:1997qn}.

The quadratic fluctuations of the scalar field~\eqref{eq_delta_phi_2_T_a_2} vanish, $\delta \langle \phi^2 \rangle_{T = T_U,a} = 0$, at the Unruh temperature~\eqref{eq_T_U}. Below, we will consider the accelerations of the thermal medium with a temperature that is higher than the Unruh temperature, $T \geqslant T_U$. The vanishing of quadratic fluctuations at temperature $T = T_U$ is in line with the Unruh-Weiss observation~\cite{Unruh:1983ac}, which states that all observables of the medium uniformly accelerating with $T = T_U$ correspond to the same observables in the Minkowski vacuum.

\vskip 1mm
\paragraph*{\bf Dirac fermions.} Now, consider Dirac fermions of the mass $M$ that are described by the Lagrangian:
\begin{align}
	{\cal L}_\psi = {\bar \psi} (i \gamma^\mu_E \partial_\mu + M) \psi\,.
\end{align}
Here, we work already in Euclidean spacetime after having performed the Wick transformation. Consequently, $\gamma^\mu$ are the Euclidean Dirac matrices that satisfy the anti-commutation relation, $\{\gamma^\mu,\gamma^\nu\} = 2 \delta^{\mu\nu}$, where $\delta^{\mu\nu}$ is the Kronecker symbol. 

We are interested in the fermionic bilinear, 
\begin{align}
    & \langle {\bar \psi} \psi\rangle = - \lim_{x \to 0} {\rm tr}\, i S(x;M),
    \label{eq_psi_2} \\
    & i S(x;M) = \langle \psi(0) {\bar \psi}(x) \rangle ,
    \label{eq_S_M}    
\end{align}
where ``tr'' denotes the trace over the spinor indices and $S(x)$ is the Euclidean fermionic propagator. The minus sign in the right-hand side of Eq.~\eqref{eq_psi_2} appears due to the anti-commutativity of the fermionic operators. The fermionic bilinear evaluated at a single point~\eqref{eq_psi_2} plays a role similar to the scalar quadratic observable~\eqref{eq_phi_2}. 

\vskip 1mm
\paragraph*{---Zero temperature $(T=0)$, no acceleration $(a = 0)$.} 
The fermionic propagator in the vacuum $S_0$ can be written in the form:
\begin{align}
	S_0(x;M) = (\gamma^\mu_E \partial_\mu - i M) G_0(x;M)\,,
    \label{eq_S_via_G}
\end{align}
where $G_0$ is the Green function of a scalar field with the mass $M$:
\begin{align}
	G_0(x;M) = \frac{M}{4 \pi^2 |x|} K_1(M |x|)\,,
    \label{eq_G_M}
\end{align}
and $K_1$ is the modified Bessel function of the second kind. In the limit of a vanishing mass, the massive scalar propagator~\eqref{eq_G_M} reduces to Eq.~\eqref{eq_G_vac}: $G_0(x,0) \equiv G_0(x)$.

Substituting Eqs.~\eqref{eq_S_via_G} and \eqref{eq_G_M} in Eq.~\eqref{eq_psi_2}, one gets the leading behavior of the $T=0$ condensate:
\begin{align}
	\langle {\bar \psi} \psi\rangle_0 = - 4 M \lim_{x \to 0} G_0(x) = - \frac{M}{\pi^2} \lim_{x \to 0} \frac{1}{|x|^2}\,,
    \label{eq_psi2_0_limit}
\end{align}
where we used the tracelessness property of the Dirac matrices, ${\rm tr}\, (\gamma^\mu_E) = 0$, and disregarded non-singular $O(M^3)$ subleading contributions. We will be interested only in the leading $O(M)$ term since the dynamical mass $M$ is small near the chiral transition point in the NJL model.

\vskip 1mm
\paragraph*{---Finite temperature $(T \neq 0)$, no acceleration $(a = 0)$.} 
The KMS conditions for fermions involve the familiar identification of the points~\eqref{eq_x_n_T}. The difference with the bosonic case arises from the fact that fermions are anticommuting operators that are subjected to the Fermi-Dirac statistics. As a result, the finite-temperature propagator gets an additional sign factor in the sum over the replicas~\eqref{eq_x_n_T}:
\begin{align}
	S_T({\boldsymbol{x}}_\perp, z,\tau ) = \sum_{n \in {\mathbb Z}} (-1)^n S_0({\boldsymbol{x}}_\perp, z,\tilde{\tau}_{(n)})\,.
    \label{eq_S_T}
\end{align}
Using Eq.~\eqref{eq_S_T} along with the leading term in the vanishing point-splitting limit~\eqref{eq_psi2_0_limit}, we get the thermal contribution to the chiral condensate which has also been given in Eq.~\eqref{eq_condensate_T} of the main text:
\begin{align}
	\delta \langle {\bar \psi} \psi\rangle_T = \langle {\bar \psi} \psi\rangle_T - \langle {\bar \psi} \psi\rangle_0 = \frac{M T^2}{6}\,.
    \label{eq_delta_psi_T}
\end{align}
Here, we again omitted the terms that enter this expression with a higher power of mass $M^l$ with $l > 1$. To take the sum over $n$ in Eq.~\eqref{eq_S_T}, we used the relation:
\begin{align}
	\sum_{n \in {\mathbb Z}} \frac{(-1)^n}{{\boldsymbol{x}}^2 + (\tau + n/T)^2} 
    = \frac{T}{2 \pi x} \frac{\sinh(\pi T x) \cos(\pi T \tau)}{\cosh(2 \pi T x) - \cos (2 \pi T \tau)}\,.
\end{align}

To make a link with a standard momentum-space formalism, let us notice that at zero temperature and vanishing acceleration, the expectation value~\eqref{eq_psi_2} can also be calculated in the path integral formalism: 
\begin{align}
	\langle {\bar \psi} \psi\rangle_0 = & - \frac{\partial}{\partial M} {\rm Det}\, (i \gamma^\mu_E \partial_\mu - M) 
    \label{eq_psi2_0}\\
    = & - \int \frac{d p_4}{2 \pi} \int \frac{d^3 p}{(2 \pi)^3} \frac{4 M}{p_4^2 + E_{\bs p}^2} 
    = - \int \frac{d^3 p}{(2 \pi)^3} \frac{2 M }{E_{\bs p}}, \nonumber
\end{align}
where ``Det'' denotes the functional determinant and $E_{\bs p} = \sqrt{{\boldsymbol{p}}^2 + M^2}$ stands for the energy of a fermion with a momentum $\boldsymbol{p}$. The quantity~\eqref{eq_psi2_0} is expectedly divergent in the ultraviolet limit, similarly to its scalar counterpart~\eqref{eq_phi_2}. The finite-temperature expression can be evaluated by passing from the integral over the time component of momentum to the sum over the fermionic Matsubara frequencies:
\begin{align}
	\int \frac{d p_4}{2 \pi} f(p_4) \to \sum_{n \in {\mathbb{Z}}} f\Bigl(2 \pi \bigl(n + 1/2\bigr) T\Bigr)\,.
\end{align}
Applying this transformation to the thermal correction to the chiral condensate via Eq.~\eqref{eq_psi2_0}, one expectedly gets Eq.~\eqref{eq_condensate_T}.

\vskip 1mm
\paragraph*{---Finite temperature $(T {\neq} 0)$ and acceleration $(a {\neq} 0)$.} 
Let us now consider the effect of acceleration on the chiral condensate. The KMS identification path~\eqref{eq_x_n_T_a} has to be supplemented with the Fermi-Dirac statistics and the spinor properties of the Dirac particle. The resulting propagator in the coordinate space reads as~\cite{Birrell1984, Ambrus2021}:
\begin{align}
	 S_{T,a}({\boldsymbol{x}}_\perp, z, \tau) &= \sum_{n \in {\mathbb Z}} (-1)^n e^{-n \alpha S^{0z}} S_0({\boldsymbol{x}}_\perp, z, \tau_{(n)}),
    \label{eq_S_T_a}
\end{align}
where we notice the spin-connection factor,
\begin{align}
	e^{-n \alpha S^{0z}} = \cos \frac{n \alpha}{2} - i \gamma^0 \gamma^3 \sin \frac{n \alpha}{2}\,, 
\end{align}
which depends on the thermal acceleration~\eqref{eq_alpha}. This factor appears in addition to the sign term in the standard finite-temperature propagator~\eqref{eq_S_T}.

The calculation of the chiral condensate can be performed in the coordinate space similarly to the evaluation of the quadratic fluctuations of the scalar field~\eqref{eq_delta_phi_2_T_a_sum}. Substituting the thermal propagator of the accelerating fermionic gas~\eqref{eq_S_T_a} to Eq.~\eqref{eq_psi_2}, subtracting the vacuum part and performing the $x \to 0$ limit, we arrive at the following formula:
\begin{align}
	\delta \langle {\bar \psi} \psi\rangle_{T,a} = \langle {\bar \psi} \psi\rangle_{T,a} - \langle {\bar \psi} \psi\rangle_0 = \frac{M T^2}{4 \pi^2} S_{F,2}(i \alpha)\,,
    \label{eq_delta_psi_T_a}
\end{align}
where, in notations of Ref.~\cite{Palermo:2021hlf},
\begin{align}
	S_{F,2}(\phi) = \sum_{n=1}^\infty (-1)^{n+1} \frac{\phi^2 \sinh(n \phi)}{\sinh^3(n \phi/2)}\,.
\end{align}
The analytically distilled value of this polynomial gives us (see Appendix~B of Ref.~\cite{Palermo:2021hlf}):
\begin{align}
	S_{F,2}(\phi) = \frac{1}{6} \Bigl(\phi^2 + (2\pi)^2 \Bigr)\,,
\end{align}
which, upon substitution to Eq.~\eqref{eq_delta_psi_T_a}, leads us to Eq.~\eqref{eq_condensate_T_a} of the main text:
\begin{align}
	\delta \langle {\bar \psi} \psi\rangle_{T,a} = \frac{M}{6} \biggl[T^2 - \Bigl(\frac{a}{2\pi}\Bigr)^2 \biggr]\,.
    \label{eq_delta_psi_T_a_final}
\end{align}
At zero acceleration, $a = 0$, the above formula reduces to the standard thermal correction to the fermionic bilinear of the non-accelerating fermionic gas~\eqref{eq_delta_psi_T}. The thermal contribution to the fermionic condensate~\eqref{eq_delta_psi_T_a_final} vanishes at the Unruh temperature~\eqref{eq_T_U}: $\delta \langle {\bar \psi} \psi\rangle_{T = T_U,a} = 0$, being consistent with the Unruh-Weiss observation~\cite{Unruh:1983ac}.

%\bibliography{njl}

\end{document}